\documentclass{article}
\usepackage[utf8]{inputenc}
\usepackage{amsmath}
\usepackage{amsthm, amssymb}
\usepackage{shadethm}
\usepackage[nothm]{thmbox}
\usepackage{graphicx}
\usepackage{thmtools}
\usepackage{subcaption}
\usepackage{xcolor}
\usepackage{cleveref}
\usepackage{algpseudocode, algorithm, algorithmicx}
\usepackage{psfrag}
\usepackage{pstool}
\usepackage{epstopdf}
\usepackage[affil-it]{authblk}
\usepackage{geometry}
\usepackage{mathtools}
\usepackage{tikz}
\usepackage{bbm}
\usepackage{bm}
\usepackage[normalem]{ulem}
\usepackage{cancel}
\usetikzlibrary{shapes,arrows,3d}
\definecolor{tumor}{HTML}{E37222}
\definecolor{tumred}{RGB}{205,32,44}
\definecolor{tumblue}{RGB}{00,115,207}
\definecolor{tumgreen}{RGB}{162,173,000}
\newcommand{\red}[1]{{\color{tumred} #1}}

\usepackage{pgfplots}
\pgfplotsset{compat=default}
\graphicspath{{graphics/}}
\usepackage{caption}
\usepackage{array}
\allowdisplaybreaks

\newcolumntype{L}{>{$}l<{$}}

\newtheorem{theorem}{Theorem}

\newtheorem{lemma}[theorem]{Lemma}
\newtheorem{definition}[theorem]{Definition}
\newtheorem{remark}[theorem]{Remark}
\newenvironment{Proof}[1][]{\proof[#1]\normalsize}{\endproof}

\newcommand{\supp}{\mathrm{supp}}
\newcommand{\sign}{\mathrm{sign}}

\renewcommand{\vec}{\mathrm{vec}}
\newcommand{\id}{\mathrm{Id}}

\newcommand{\B}{\mathcal{B}}

\renewcommand{\P}{\mathbb{P}}

\newcommand{\R}{\mathbb{R}}

\newcommand{\set}[1]{\mathcal{#1}}

\newcommand{\E}[2][]{\textnormal{\;\textsf{E}}_{#1}\!\left[#2\right]}

\renewcommand{\P}[2][]{\textnormal{\;\textsf{P}}_{#1}\!\left[#2\right]}
\newcommand{\reals}{\mathbb{R}}
\newcommand{\x}{\mathbf{x}}
\newcommand{\X}{\mathbf{X}}

\newcommand{\y}{\mathbf{y}}
\newcommand{\Y}{\mathbf{Y}}

\renewcommand{\a}{{\mathbf{a}}}
\renewcommand{\b}{{\mathbf{b}}}
\newcommand{\al}{\mathbf{a}^{\!(l)}}
\newcommand{\bl}{\mathbf{b}^{\!(l)}}

\newcommand{\tl}{\boldsymbol{\vartheta}^{(l)}}
\newcommand{\thl}{\hat{\boldsymbol{\vartheta}}^{(l)}}

\newcommand{\A}{{\mathbf{A}}}
\newcommand{\Al}[1]{\mathbf{A}^{\hspace{-.2em}(#1)}}
\newcommand{\Bmat}{{\mathbf{B}}}

\newcommand{\eps}{\varepsilon}
\renewcommand{\log}{\ln}

\newcommand{\0}{\mathbf{0}}
\newcommand{\eqo}[1]{\overset{\textnormal{#1}}{=}}

\newcommand{\leqo}[1]{\overset{\textnormal{#1}}{\leq}}

\newcommand{\g}{\mathbf{g}}
\newcommand{\ip}[2]{\left\langle #1, #2 \right\rangle}
\newcommand{\pnorm}[2]{\left\| #1 \right\|_{#2}}
\newcommand{\z}{\mathbf{z}}
\newcommand{\Z}{\mathbf{Z}}
\renewcommand{\a}{{\mathbf{a}}}

\renewcommand{\red}[1]{#1}

\renewcommand{\log}{\ln}

\newcommand{\comm}[1]{{}}

\title{Analysis of Hard-Thresholding for Distributed Compressed Sensing with One-Bit Measurements}
\author[1]{Johannes Maly} 
\author[2]{Lars Palzer\thanks{The authors gratefully acknowledge the support by the DFG project "Information Theory and Recovery Algorithms for Quantized and Distributed Compressed Sensing" in context of SPP 1798.}}
\affil[1]{Department
of Mathematics, Technical University of Munich, Germany}
\affil[2]{Department
of Electrical and Computer Engineering, Technical University of Munich, Germany}

\begin{document}

\maketitle

\begin{abstract}
    A simple hard-thresholding operation is shown to be able to uniformly recover $L$ signals $\x_1,...,\x_L \in \R^n$ that share a common support of size $s$ from $m = \set{O}(s)$ one-bit measurements per signal if $L \ge \log(en/s)$. This result improves the single signal recovery bounds with $m = \set{O}(s\log(en/s))$ measurements in the sense that asymptotically fewer measurements per non-zero entry are needed. Numerical evidence supports the theoretical considerations.
\end{abstract}


\section{Introduction} \label{sec:Introduction}
The field of compressed sensing emerged from the following basic question: Given some unknown and high-dimensional signal $\x \in \R^n$, 
what is the smallest number $m$ of linear measurements
\begin{align} \label{eq:CS}
    \y = \A\x
\end{align}
needed to uniquely determine $\x$, where $\A \in \R^{m\times n}$. From basic linear algebra we require $m \ge n$ in general.

Now suppose that $\x$ is $s$-sparse, i.e., $\x$ has at most $s \le n$ entries that are not equal to zero, and let $\supp(\x) = \{ i \in [n] \colon x_i \neq 0 \} \subset [n] := \{1,\dots,n\}$ be the support of $\x$. Knowing $\supp(\x)$ now $m \ge s$ measurements suffice to uniquely identify $\x$ from $\y$ (assuming that every choice of $s$ different colums of $\A$ is linearly independent). If $s \ll n$, this yields a considerable improvement in the number of measurements. In practice, however, the support of $\x$ is unknown. The seminal works \red{\cite{candes2006stable}, \cite{candes2006robust}, \cite{donoho2006compressed}} lay foundations of compressed sensing by guaranteeing even then unique identification of all $s$-sparse $\x$ from $m < n$
measurements defined in \eqref{eq:CS} if $\A$ is suitably chosen. Moreover, efficient recovery is possible by means of convex optimization. A sufficient condition for $\A$ to allow this is the so-called \textit{restricted isometry property} (RIP). A linear operator $\A$ satisfies the RIP of order $s$ with RIP-constant $\delta \in (0,1)$ if
\begin{align} \label{eq:CSrip}
    (1-\delta) \| \x \|_2 \le \| \A\x \|_2 \le (1+\delta) \| \x \|_2
\end{align}
for all $s$-sparse $\x$, i.e., $\A$ embeds the set of $s$-sparse $n$-dimensional vectors almost isometrically into $\R^m$ (see \red{\cite{candes2006stable}, \cite{candes2006near}}). Though no deterministic construction of RIP matrices has been discovered so far for less than $m = \set{O}(s^2)$ measurements (cf.\ \red{\cite[Ch 6]{foucart:2013}}), several classes of randomly generated matrices satisfy the RIP with exceedingly high probability if
\begin{align} \label{eq:CSmeasurements}
    m \ge Cs \log\left(\frac{en}{s}\right)
\end{align}
where $C > 0$ is a constant independent of $s,m,$ and $n$ (see \red{\cite{candes2006near}, \cite{rudelson2008sparse}}). Hence, up to the log-factor, $\mathcal{O}(s)$ measurements suffice to capture all information in the $n$-dimensional signal $\x$. Moreover, the bound~\eqref{eq:CSmeasurements} is robust to noise on the measurements.\\

However, there are more difficulties to overcome than just noise. In particular, real-valued measurements $y_i \in \R$ cannot not be stored with infinite precision. The idealistic measurement model  presented in \eqref{eq:CS} should be extended by a quantizer $Q$ that maps the real-valued measurement vector $\A\x$ to a finite alphabet. The extreme case is to choose $Q$ as the sign function acting componentwise on $\A\x$ leading to the \textit{one-bit compressed sensing} model first studied in \red{\cite{boufounos:2008}}
\begin{align} \label{eq:CSquantized}
    \y = \sign(\A\x)
\end{align}
i.e., $y_i$ is $1$ if $\langle \a_i,\x \rangle \ge 0$ and $-1$ if $\langle \a_i,\x \rangle < 0$, where $\a_i$ is the $i$-th row of $\A$. From a geometric point of view, this single bit expresses on which side of the hyperplane $H_{\a_i}$ (defined by the normal vector $\a_i$) the signal $\x$ lies. Note that the model~\eqref{eq:CSquantized} is blind to scaling and we can only hope to approximate the direction of $\x$ (this issue can be tackled by, e.g., adding a random \textit{dither} to the measurements before quantization and thus shifting the hyperplanes away from the origin, see \red{\cite{knudson2016one}, \cite{baraniuk2017exponential}}).

It turns out that for one-bit quantization, the bound in \eqref{eq:CSmeasurements} still provides a sufficient number of measurements to approximate all $s$-sparse $\x$ of unit norm by convex optimization, see \red{\cite{plan:2013}}, or greedy approaches like a single hard-thresholding step, see \red{\cite{jacques:2013}}. 
Here, approximating means the following: One cannot recover $\x$ from $\y$ exactly but one can bound the worst case error (in, e.g., $\ell_2$ norm) of certain reconstruction algorithms. The difference between the required measurements for \eqref{eq:CS} and \eqref{eq:CSquantized} lies in approximation quality: the expected worst-case error is much better (possibly zero) in unquantized compressed sensing. Nevertheless, one-bit sensing is of great interest for applications because single bit measurement devices are cheap to produce.



Another method to reduce the number of measurements is \textit{distributed compressed sensing} or joint recovery (see \red{\cite{baron2009distributed}, \cite{duarte2005joint}}). The idea is that the log-factor in \eqref{eq:CSmeasurements} is caused by not knowing the exact support of $\x$. If one has to recover several signals $\x_1,...,\x_L$, $L \in \mathbb{N}$, sharing a common support, it might be possible to reduce the number of measurements per signal from $\set{O}(s\log(en/s))$ to $\set{O}(s)$ by profiting from the joint structure. In theory the improvement seems small, but in practice it can make a notable difference (cf.\ \red{\cite{sundman2013methods}}). Moreover, the joint support assumption might appear naturally. \red{For example, a signal that is sparse in the Fourier basis may be measured at different locations, which leads to different attenuations and phase shifts at every node. This can be exploited in imaging applications such as MRI~\cite{wu:2014}. Another prominent application is MIMO communications~\cite{rao:2014}.} 

There are two popular settings for joint recovery from compressed measurements. The first model is called {\em Multiple Measurement Vectors (MMV)}. All signals are measured by the same measurement matrix $\A \in \R^{m\times n}$ (resp. the same sensor) and the model in \eqref{eq:CS} becomes
\begin{align} \label{eq:DCSoneMatrix}
    \Y = \A\X
\end{align}
where $\X \in \R^{n\times L}$ and $\Y \in \R^{m\times L}$ are matrices containing the signals and their corresponding measurement vectors as columns. As shown in \red{\cite{eldar:2010}}, for this model one can improve only the average performance as compared to single vector compressed sensing. The worst-case analysis shows no improvement. 

The second model considers distinct measurement matrices $\Al{1},...,\Al{L} \in \R^{m\times n}$ (resp. distinct sensors) for each signal $\x_l \in \R^n$, $l \in [L]$. Hence, there is a seperate measurement process of type \eqref{eq:CSmeasurements} for each $l \in [L]$ yielding $L$ different $\y_l \in \R^m$. We have
\begin{align} \label{eq:DCSmultMatrices}
    \vec(\Y) = \A \cdot \vec(\X)
\end{align}
where $\A \in \R^{mL \times nL}$ is block diagonal and built from the blocks $\Al{l}$, and $\vec(\cdot)$ denotes the vectorization of a matrix. The authors of \red{\cite{eldar:2009}} guarantee recovery of jointly sparse signal ensembles $\X$ from measurements of type \eqref{eq:DCSmultMatrices} via $\ell_{2,1}$-minimization provided $\A$ satisfies a certain block RIP. A direct connection between the number of measurements to guarantee block RIPs for random matrices and properties of the signal ensembles $\X$ is presented in \red{\cite{eftekhari2015restricted}}. In particular, the authors show that one can profit from joint structure if the information in $\X$ is spread among multiple signals $\x_l$. For instance, if all $\x_l$ but one are zero one will need $m = \set{O}(s\log(en/s))$ measurements per signal rendering joint recovery useless. Hence, to obtain meaningful recovery guarantees for distributed compressed sensing one needs assumptions beyond a joint support set (see also Remark \ref{rem:singleHT} below).\\

Both extensions of the classical compressed sensing model \eqref{eq:CS}, namely, one-bit compressed sensing and distributed compressed sensing, are useful in practice. One might thus try to combine both approaches to reduce the number of measurements in one-bit sensing. The papers \red{\cite{tian2014distributed}, \cite{kafle2016decentralized}, \cite{gupta2015joint}} show promising numerical results, but they do not provide theoretical justification for the improvements.


\subsection{Contribution}

We provide uniform approximation guarantees for distributed compressed sensing from one-bit measurements quantifying the influence of the size $L$ of signal ensembles $\X$ on the required number of measurements per signal $m$. Our analysis considers the second model above, i.e., distinct measurement matrices $\Al{1},...,\Al{L}$ corresponding to distinct sensors. In particular, we show that if the entries of all $\Al{l}$ are drawn as independent and identically distributed (iid) Gaussian random variables, then the matrix $\A$ will satisfy an $\ell_1/\ell_{2,1}$-RIP on a suitable set of jointly sparse signal ensembles with high probability. We adapt the ideas of \red{\cite{foucart:2016}} to deduce a uniform error bound for recovering appropriate signal ensembles $\X$ from their one-bit measurements $\Y$ by applying one simple hard-thresholding step to $\A^T \vec(\Y)$. We find that $mL \ge Cs (\log(en/s) + L)$ measurements suffice to well-approximate $\X$ with high probability which means, for $L \simeq \log(en/s)$, $\set{O}(s)$ measurements per single signal. This improves the classical compressed sensing results for Gaussian measurements of $\set{O}(s\log(en/s))$ (cf.\ \red{\cite{foucart:2013}}). Moreover, we provide numerical evidence matching anterior experimental results in \red{\cite{tian2014distributed}, \cite{kafle2016decentralized}, \cite{gupta2015joint}}.\\

We remark that we recently became aware of the related work \cite{genzel2017recovering}. The authors examine recovery of structured data from superimposed non-linear measurements by group-LASSO, i.e., $\ell_{2,1}$-minimization. Interpreting their results from our viewpoint, the techniques might provide similar (in number of necessary measurements) non-uniform approximation guarantees for $\ell_{2,1}$ minimization in our setting. This is, however, not obvious since their setting focuses on recovering one single signal from superimposed distributed measurements while we consider different signals and full knowledge of all distributed measurements. The different analysis methods of both works are complementary, and both serve to better understand distributed compressed sensing with one-bit measurements. Our approach has the advantage of brief and elementary proofs, while the approach of \cite{genzel2017recovering} might give improved error bounds using more sophisticated tools.


\subsection{Outline}

The paper is organized as follows. Section \ref{sec:ProblemSetup} introduces our problem in detail, Section \ref{sec:MainResults} presents our main results, and Section \ref{sec:Proofs} gives proofs. Section \ref{sec:NumericalExperiments} supports the validity of the theory by numerical experiments, and Section \ref{sec:Conclusion} concludes with a brief summary and outlook on future work.


\subsection{Notation}

We abbreviate $[n] := \{1,...,n\}$. We use bold font for vectors and matrices and regular font for scalars. Matrices are denoted by capital letters to distinguish them from vectors. $\id_n$ represents the $n$-dimensional identity matrix. The vector $\z \in \R^{n}$ has entries $z_i \in \R$ and the matrix $\Z \in \R^{n_1 \times n_2}$ has entries $Z_{i,j}, i \in [n_1], j \in [n_2]$. We write the $p$-norms of vectors as $\| \cdot \|_p$. Note that the Frobenius norm of matrices $\| \cdot \|_F$ corresponds to the $\ell_2$-norm of the vectorization. We use the matrix norm $\pnorm{\Z}{2,1}$ which is the sum of the $\ell_2$-norms of the columns of $\Z$ and, by abuse of notation, write $\pnorm{\z}{2,1} = \pnorm{\Z}{2,1}$ if $\z = \vec(\Z)$ is the vectorized representation of a matrix $\Z$. We denote the Euclidean ball of radius $1$ centered at $\0$ by $\set{B}(\0,1)$. The scalar sign-function is defined as $\sign(z) = -1$ if $z < 0$ and $\sign(z) = 1$ if $z \ge 0$, and acts componentwise if applied to vectors. We will use $\lesssim$ and $\gtrsim$ to denote $\le$ resp. $\ge$ up to an absolute multiplicative constant. The Gaussian distribution with mean $\boldsymbol{\mu}$ and covariance matrix $\Sigma$ is denoted by $\set{N}(\boldsymbol{\mu},\Sigma)$. Probabilities are denoted by $\P{\cdot}$ and expectations by $\E{\cdot}$. Random and deterministic quantities will be denoted using the same font as it should be clear from the context which quantities are random.


\section{Problem Setup} \label{sec:ProblemSetup}
Suppose we are given one-bit measurements $\Y \in \R^{m\times L}$ obtained from $L$ signals $\x_l \in \R^n$, $l \in [L]$, that form the columns of a matrix $\X \in \R^{n\times L}$. For simplicity we write $\x = \vec(\X) = (\x_1^T,...,\x_L^T)^T$ and $\y = \vec(\Y) = (\y_1^T,...,\y_L^T)^T$. The linear measurement process can then be described by
\begin{align} \label{eq:measurements}
	\y = \sign\bigl((\theta \A) \x\bigr)
\end{align}
where $\theta > 0$ is an appropriate scaling to be determined later (in fact, by scaling blindness of $\sign$ the explicit choice of $\theta$ has no influence on the measurements) and $\A \in \R^{Lm \times Ln}$ is a measurement matrix of the following form: $\A$ is block diagonal and built from the submatrices $\Al{l} \in \R^{m\times n}$, $l \in [L]$, which have iid Gaussian entries $\set{N}(0,1)$, i.e., we have
\begin{align} \label{eq:defA}
	\theta \A = \theta \begin{pmatrix}
	\multicolumn{1}{c|}{\Al{1}} & & \\ \cline{1-1}
	 & \ddots & \\ \cline{3-3} 
	& & \multicolumn{1}{|c}{\Al{L} } 
	\end{pmatrix}. 
\end{align}
We denote the $i$-th column of $(\Al{l})^T$ by $\al_i$, i.e., $\al_i$ is the transposed $i$-th row of $\Al{l}$. We aim to approximate $\x$ by a single hard-thresholding step (cf.\ \red{\cite{foucart:2016}})
\begin{align} \label{eq:singleHT}
    \hat{\x} = \mathbb{H}_s\left((\theta\A)^T\y\right)
\end{align}
where the hard-thresholding operator $\mathbb{H}_s (\z) = \mathbb{H}_s (\vec(\Z)) = \vec(\mathbb{H}_s (\Z))$ keeps only the $s$ rows of largest $\ell_2$-norm, for all $\z = \vec(\Z) \in \R^{Ln}$ with $\Z \in \R^{n\times L}$. We will see that this simple procedure leads to near-optimal approximation guarantees for signal ensembles $\X$ whose signals $\x_l$ share a common support and the same magnitude in $\ell_2$-norm. We denote the support of a signal ensemble $\Z \in \R^{n\times L}$, i.e., the set of non-zero rows of $\Z$, by $\supp(\Z) \subset [n]$. We define the set $\set{S}_{s,L}$ of admissible signal ensembles
\begin{align} \label{eq:S_sL}
\set{S}_{s,L} = \biggl\{ \z = \vec{(\Z)} \colon \Z = \begin{pmatrix}
| & & | \\
\z_1 & \cdots & \z_L \\
| & & |
\end{pmatrix}
\in\reals^{n\times L}, |\supp(\Z)| \leq s, \|\z_l\|_2 = \pnorm{\z}{2}/\sqrt{L}\biggr\}. 
\end{align}
As the simple sign-bit measurements \eqref{eq:measurements} are invariant under scaling of the signals and, hence, dismiss any information on signal magnitudes, all we can hope for is approximating the directions of the individual signals. Hence, we can restrict the $\x_l$ to have constant norm without loss of generality. Consequently, whenever we use the terms "approximation of signals" or "recovery of signals" we implicitly mean "approximation/recovery of each signal up to the scaling" and restrict the results to signals of fixed norm.



\section{Main Results} \label{sec:MainResults}
We show that Gaussian measurements of the form \eqref{eq:defA} fulfill under suitable scaling with high probability an $\ell_1/\ell_{2,1}$-\textit{restricted isometry property} ($\ell_1/\ell_{2,1}$-RIP) on
\begin{align} \label{eq:K_sL}
\set{K}_{s,L} = \biggl\{ \z = \vec{(\Z)} \colon \Z \in\reals^{n\times L}, |\supp(\Z)| \leq s \biggr\}
\end{align}
(a relaxation of $\set{S}_{s,L}$) if $mL \gtrsim s(\log(en/s)+L)$ and that a fixed signal ensemble $\X \in \set{S}_{s,L}$ can be well approximated from $mL \gtrsim s(\log(en/s)+L)$ one-bit measurements following the aforementioned model \eqref{eq:measurements}. Proofs can be found in Section \ref{sec:Proofs}. Let us first define what we mean by $\ell_1/\ell_{2,1}$-RIP.
\begin{definition}[$\ell_1/\ell_{2,1}$-RIP]
    A matrix $\Bmat \in \R^{Lm \times Ln}$ satisfies the $\ell_1/\ell_{2,1}$-RIP on $\set{K}_{s,L}$ with RIP-constant $\delta \in (0,1)$ if
    \begin{align} \label{eq:RIP}
        \frac{\| \z \|_{2,1}}{\sqrt{L}} - \delta \pnorm{\z}{2} \le \| \Bmat\z \|_1 \le \frac{\| \z \|_{2,1}}{\sqrt{L}} + \delta \pnorm{\z}{2}
    \end{align}
    for all $\z \in \set{K}_{s,L}$.
\end{definition}
The following lemma provides a sufficient number of measurements for $\theta \A$ in \eqref{eq:defA} to fulfill the above introduced $\ell_1/\ell_{2,1}$-RIP. Its proof is inspired by \cite[Cor.\ 2.3]{plan:2014}.
\begin{lemma}[$\ell_1/\ell_{2,1}$-RIP] \label{lem:RIP}
    For $\theta = \sqrt{\pi/(2Lm^2)}$ and $mL \gtrsim \delta^{-2} s(\log(en/s) + L)$ the operator $\theta \A$ as defined in \eqref{eq:defA} has the $\ell_1/\ell_{2,1}$-RIP on $\set{K}_{s,L}$ with RIP-constant $\delta$ with probability at least\\ $1-2\exp \left( -\delta^2 mL/ (4\pi) \right)$.
\end{lemma}
\begin{remark} \label{rem:RIP}
    For $L=1$ this result agrees with known bounds on the sufficient number of measurements to have $\ell_1/\ell_2$-RIPs for random Gaussian matrices with high probability, namely, $m \gtrsim s\log(en/s)$. If $L \ge \log(en/s)$, we have $(\log(en/s)+L)/L \le 2$ and, hence, Lemma \ref{lem:RIP} requires $m \gtrsim \delta^{-2}s$ for an RIP on signal ensembles in $\set{K}_{s,L}$, i.e., only $\mathcal{O}(s)$ measurements per signal. 
    \\ 
    In \cite{eftekhari2015restricted} the authors examined how many measurements suffice for random Gaussian block matrices $\A$ to satisfy classical $\ell_2$-RIPs depending on how the information of sparse signals is distributed on the different blocks of $\A$. Lemma \ref{lem:RIP} extends their result to $\ell_1/\ell_{2,1}$-RIPs in the special case that all signals have the same support.\\
    As $\pnorm{\z}{2,1} \le \sqrt{L} \pnorm{\z}{2}$, the upper bound in \eqref{eq:RIP} can be replaced by $(1+\delta) \pnorm{\z}{2}$. Moreover, if restricted to $\set{S}_{s,L}$ the $\ell_1/\ell_{2,1}$-RIP in \eqref{eq:RIP} becomes a full $\ell_1/\ell_2$-RIP, i.e.,
    \begin{align*}
        (1-\delta) \pnorm{\x}{2} \le \pnorm{\Bmat\x}{1} \le (1+\delta) \pnorm{\x}{2}
    \end{align*}
    as in this case $\pnorm{\x}{2,1} = \sqrt{L} \pnorm{\x}{2}$. This observation suggests that the signal ensemble model $\set{S}_{s,L}$ is well-chosen as the ensembles in $\set{S}_{s,L}$ when multiplied to block-diagonal Gaussian measurement matrices (induced by the distributed setting) behave like single vectors multiplied to dense Gaussian measurement matrices.
\end{remark}
The next theorem is the main result of this paper. It guarantees uniform recovery of all signal ensembles $\x \in \set{S}_{s,L}$ by a simple hard-thresholding step. This result generalizes \cite[Thm.\ 8]{foucart:2016} to joint recovery of signals sharing a common support.
\begin{theorem} \label{thm:singleHT}
    Let $n,m,s >0$, and let $\A$ be a random $Lm\times Ln$ matrix as defined in~\eqref{eq:defA}. Set
    \begin{align} \label{eq:MeasBound}
        mL \gtrsim \delta^{-2} s (\log(en/s) + L)
    \end{align}
    Then with probability at least $1-2\exp \left( -\delta^2 mL/ (4\pi) \right)$ (over the entries of $\A$), we have
    for all $\x \in \set{S}_{s,L}$ with $\pnorm{\x}{2} = 1$ that
    \begin{align} \label{eq:ErrorBound}
        \pnorm{\x - \hat{\x}}{2} \lesssim \sqrt{\delta}
    \end{align}
    where $\hat{\x}$ is defined in \eqref{eq:singleHT}.
\end{theorem}
\begin{remark} \label{rem:singleHT}
    We would like to stress some points:\\
    $(i)$ As already mentioned in Remark \ref{rem:RIP}, the required number of measurements per signal does not depend on $s \log(n/s)$ if $L \ge \log(n/s)$ but only on $s$, i.e., when recovering several signals that share a common support from sign-measurements collected independently for each single signal, one can significantly reduce the necessary number of measurements by using the support structure.\\
    $(ii)$ For unit norm signals $\| \x_l \| = 1$ the error bound in \eqref{eq:ErrorBound} becomes
    \begin{align*}
        \pnorm{\x - \hat{\x}}{2} \lesssim \sqrt{L\delta}
    \end{align*}
    i.e., the error per single signal $\x_l$ is only less than $\sqrt{\delta}$ on average. In the worst case this error concentrates on one signal. However, if the signals all are dense on shared support set $\set{T} \subset [n]$, the support will be recovered even in this case as a worse error on one signal implies less error on the remaining signals. It is not surprising that a dense support of all signals is necessary to profit from joint recovery. If only one signal has dense support while the rest contains mostly zeros on $\set{T}$, then most of the signals do not carry helpful support information, i.e., joint recovery cannot be expected to improve performance.\\
    $(iii)$ At first sight, the proof of Theorem \ref{thm:singleHT} hardly differs from the one of \cite[Thm.\ 8]{foucart:2016}. One first proves an $\ell_1/\ell_2$-RIP for $\A$ and then concludes by a simple computation. However, the model selection $\set{S}_{s,L}$ is crucial and one has to treat the matrix $\A$ as a whole to reach the sample complexity in \eqref{eq:MeasBound}.  Consider the following naive approach: If we have $m \gtrsim \delta^{-2} s \log(en/s)$ for some $\delta > 0$, we get for each $l \in [L]$ and Gaussian $\Al{l} \in \reals^{m\times n}$ that with probability exceeding $1-C\exp(-c\delta^2m)$
    \begin{align} \label{eq:singleRIP}
        (1-\delta) \pnorm{\z}{2} \le \frac{\sqrt{2}}{m\sqrt{\pi}} \pnorm{\Al{l}\z}{1} \le (1+\delta) \pnorm{\z}{2}
    \end{align}
    for all $s$-sparse $\z \in \reals^n$~(see \cite{schechtman2006two}). By a union bound and summation over \eqref{eq:singleRIP} for $l \in [L]$ one obtains with probability at least $1 - C\exp(-c\delta^2m + \log(L))$ that
    \begin{align*}
        (1-\delta) \pnorm{\x}{2} \le \pnorm{(\theta \A)\x}{1} \le (1+\delta) \pnorm{\x}{2}
    \end{align*}
    for all $\x \in \set{S}_{s,L}$ and $\theta = \sqrt{2}/(m\sqrt{\pi L})$. Choosing $\delta' = \sqrt{L}\delta$ (to obtain comparable probabilities of success) shows that this leads to a worse sample complexity than \eqref{eq:MeasBound}. \\
    $(iv)$ The proof of Theorem \ref{thm:singleHT} relies on the assumption that $x \in \set{S}_{s,L}$. As mentioned in Remark \ref{rem:RIP} this assumption corresponds to the equivalence of $\ell_1/\ell_{2,1}$-RIP and $\ell_1/\ell_2$-RIP on $\set{S}_{s,L}$. It is possible to relax the restriction a little. To this end, define for $\eps \in (0,1)$ the set
    \begin{align}
    \set{S}_\eps = \biggl\{ \z = \vec{(\Z)} : \Z\in\reals^{N\times L}, \; \supp(\Z)\leq s, \; \|\z_l\|_2 \in \left[ \frac{1-\eps}{\sqrt{L}} \pnorm{\z}{2}, \frac{1+\eps}{\sqrt{L}} \pnorm{\z}{2} \right] \biggr\}.
    \end{align}
    of signal ensembles which differ in norm by a bounded perturbation. Assume that a matrix $\Bmat$ satisfies the $\ell_1/\ell_{2,1}$-RIP on $\set{K}_{s,L}$ with RIP-constant $\delta > 0$. By noting that $\pnorm{\x}{2,1} \in [1-\eps,1+\eps]\sqrt{L} \pnorm{\x}{2}$ if $\x \in \set{S}_\eps$, this implies
    \begin{align} \label{eq:deRIP}
        (1-\delta)(1-\eps) \pnorm{\x}{2} \le \pnorm{\Bmat\x}{1} \le (1+\delta)(1+\eps) \pnorm{\x}{2}.
    \end{align}
    To rewrite \eqref{eq:deRIP} as an $\ell_1/\ell_2$-RIP on $\set{S}_\eps$ for some $\delta' \in (0,1)$, i.e.,
    \begin{align} \label{eq:perturbedRIP}
        (1-\delta') \pnorm{\x}{2} \le \pnorm{\Bmat\x}{1} \le (1+\delta') \pnorm{\x}{2}
    \end{align}
    for all $\x \in \set{S}_\eps$, it would suffice that
    \begin{align*}
        (1-\delta') \le (1-\delta)(1-\eps)
    \end{align*}
    which is equivalent to
    \begin{align*}
        \eps \le \frac{\delta' - \delta}{1-\delta}.
    \end{align*}
    As the right-hand side is only positive for $\delta < \delta'$ and a decreasing function in $\delta$ for $0 \le \delta < \delta'$ we can upper bound it by $\delta'$. Hence, the more general $\ell_1/\ell_{2,1}$-RIP only becomes an $\ell_1/\ell_2$-RIP on $\set{S}_\eps$ for $\eps \le \delta'$ meaning that only small perturbations $\eps$ are possible if the approximation error in \eqref{eq:ErrorBound} shall be small. However, the assumption of all signals $\x_l$ sharing the same norm is a mild condition in our setting as~\eqref{eq:CSquantized} is blind to scaling and norm variations in signal ensembles.
\end{remark}


\section{Proofs} \label{sec:Proofs}
We provide here the proofs of Lemma \ref{lem:RIP} and Theorem \ref{thm:singleHT}. As both proofs rely on an improved understanding of $\set{K}_{s,L}$ defined in \eqref{eq:K_sL}, we start by analizing this set in Section \ref{sec:PropertiesOfS}. The proof of Lemma \ref{lem:RIP} can be found in Section \ref{sec:LemmaRIP} and the proof of Theorem \ref{thm:singleHT} is presented in Section \ref{sec:TheoremHT}.


\subsection{Properties of $\set{K}_{s,L}$} \label{sec:PropertiesOfS}

An important measure of complexity for subsets of $\R^d$ is the so-called \textit{Gaussian width}. This quantity generalizes the notion of linear dimension to arbitrary sets and is a useful tool for estimating the sampling requirements of signal sets in compressed sensing.
\begin{definition}[{Gaussian width \cite[Eq. (1.2)]{plan:2014}}]
    The Gaussian width of $\set{K} \subset \R^d$ is defined as
    \begin{align*}
        w(\set{K}) = \E{\sup_{\z \in \set{K}} |\langle \g,\z \rangle|}
    \end{align*}
    where $\g \sim \set{N}(\0,\id_d)$ is a random vector with iid Gaussian entries.
\end{definition}
\begin{remark}
    Examples illustrating the relation between Gaussian width and set complexity are as follows \cite{plan:2013a}:
    \begin{enumerate}
        \item[$(i)$] For $\set{K} = \set{B}(\0,1) \subset \R^d$ one has $w(\set{K}) \lesssim \sqrt{d}$.
        \item[$(ii)$] If the linear dimension of a set $\set{K} \subset \set{B}(\0,1) \subset \R^d$ is $\dim(\set{K}) = k$, then $w(\set{K}) \lesssim \sqrt{k}$.
        \item[$(iii)$] Let $\Sigma_s \subset \R^d$ denote the set of $s$-sparse vectors. Then $w(\Sigma_s \cap \set{B}(\0,1)) \lesssim s\log(ed/s)$.
    \end{enumerate}
    Examples $(i)$ and $(ii)$ show that $w(\set{K})$ provides a consistent extension of the linear dimension to arbitrary sets in $\R^d$. A helpful rule of thumb is $w(\set{K})^2 \sim \dim(\set{K})$, i.e., the complexity of a set corresponds to the squared Gaussian width. However, note that contrary to $\dim(\set{K})$ the Gaussian width scales with $\sup_{\z \in \set{K}} \| \z \|_2$.
\end{remark}
The Gaussian width of a set $\set{K}$ is closely related to the covering number $N(\set{K},\eps)$ via Dudley's and Sudakov's inequalities (cf.\ \cite{talagrand2014upper}). The covering number $N(\set{K},\eps)$ of a set is defined as the minimal number of $\eps$-balls in $\ell_2$-norm (centered in $\set{K}$) one needs to cover $\set{K}$ completely. The cardinality of any $\eps$-net of a set $\set{K}$ provides an upper bound of $N(\set{K},\eps)$. A subset $\tilde{\set{K}} \subset \set{K}$ is called an $\eps$-net of $\set{K}$ if for any $\z \in \set{K}$ there exists $\tilde{\z} \in \tilde{\set{K}}$ with $\| \z - \tilde{\z} \|_2 \le \eps$. We obtain a bound on $w(\set{K}_{s,L} \cap \set{B}(\0,1))$ by first bounding $N(\set{K}_{s,L} \cap \set{B} (\0,1),\eps)$ in Lemma \ref{lem:CoveringNumber} and then applying Dudley's inequality in Lemma \ref{lem:GaussianWidth}.
\begin{lemma}[Covering Number of $\set{K}_{s,L} \cap \set{B} (\0,1)$] \label{lem:CoveringNumber}
    For $\eps \in (0,1)$ we have
    \begin{align*}
        \log\left( N(\set{K}_{s,L} \cap \set{B} (\0,1),\eps)\right) \le s \log\left( \frac{en}{s}\right) + sL \log\left( \frac{3}{\eps} \right).
    \end{align*}
\end{lemma}
\begin{Proof}
    As $\set{K}_{s,L} \cap \set{B} (\0,1)$ is the union of $\binom{n}{s}$ unit $\ell_2$-balls in $\R^{sL}$ embedded into $\R^{nL}$ and each unit ball can be covered by an $\eps$-net of cardinality at most $(3/\eps)^{sL}$ (see \cite[Section 3]{candes2011tight}), we know that
    \begin{align*}
        N\left(\set{K}_{s,L} \cap \set{B} (\0,1),\eps \right) \le \binom{n}{s} \left( \frac{3}{\eps} \right)^{sL} \le \left( \frac{en}{s} \right)^{s} \left( \frac{3}{\eps} \right)^{sL}. 
    \end{align*}
\end{Proof}
Lemma \ref{lem:CoveringNumber} leads to a direct bound for $w(\set{K}_{s,L} \cap \set{B}(\0,1))$.
\begin{lemma}[Gaussian width of $\set{K}_{s,L} \cap \set{B} (\0,1)$] \label{lem:GaussianWidth}
    We have
    \begin{align*}
        w(\set{K}_{s,L} \cap \set{B}(\0,1)) \lesssim \sqrt{s \left( \log \left( \frac{en}{s} \right) + L \right)}.
    \end{align*}
\end{lemma}
\begin{Proof}
    By \cite[Prop.\ 2.1]{plan:2013a} one has $w(\set{K}) =  \E{\sup_{\z \in \set{K}} \langle \g,\z \rangle }$ for an origin symmteric set $\set{K}$. Hence, we obtain
    \begin{align*}
        w(\set{K}_{s,L} \cap \set{B}(\0,1)) &\le  \E{\sup_{\z \in \set{K}_{s,L} \cap \set{B}(\0,1)} \langle \g,\z \rangle} \leqo{(i)} 24 \int_0^1 \sqrt{\log\left( N(\set{K}_{s,L} \cap \set{B}(\0,1), \eps) \right)} \; d\eps \\
        &\leqo{(ii)} 24 \sqrt{\int_0^1 1^2 \; d\eps} \cdot \sqrt{\int_0^1 \log\left( N(\set{S}_{s,L} \cap \set{B}(\0,1), \eps) \right) \; d\eps} \\
        &\leqo{(iii)} 24\sqrt{s \left( \log \left( \frac{en}{s} \right) + L (1+\log 3) \right)}
    \end{align*}
    where (i) follows from Dudley's inequality~\cite[Thm. 11.7]{ledoux:2013}, (ii) from H\"older's inequality and (iii) from Lemma \ref{lem:CoveringNumber}.
\end{Proof}


\subsection{Proof of Lemma \ref{lem:RIP}} \label{sec:LemmaRIP}

To prove the $\ell_1/\ell_{2,1}$-RIP for $\theta \A$ on the signal set $\set{K}_{s,L}$ we will restrict ourselves to $\set{K}_{s,L} \cap \mathbb{S}^{nL-1}$ where $\mathbb{S}^{nL-1}$ denotes the unit sphere in $\reals^{nL}$. It suffices to show \eqref{eq:RIP} for all $\z \in \set{K}_{s,L} \cap \mathbb{S}^{nL-1}$ as \eqref{eq:RIP} is invariant under scaling of the $\ell_2$-norm. The proof hence reduces to a direct application of the following concentration lemma which is a slightly adapted version of \cite[Lemma 2.1]{plan:2014}. For sake of completeness we report its full proof.
\begin{lemma} \label{concentrationLemma}
Consider a bounded subset $\set{K} \subset \reals^{NL}$ and let $\al_i\sim \set{N}(\0,\id_n)$, $i\in [m]$, $l\in [L]$ be independent Gaussian vectors in $\reals^n$.
Define
\begin{align}
    Z \coloneqq
    \sup_{\x\in\set{K}} \left| \sum_{i=1}^m \sum_{l=1}^L \sqrt{\frac{\pi}{2Lm^2}} \left|\ip{ \al_{i}}{\x_{l}}\right| - \frac{1}{\sqrt{L}}\pnorm{\x}{2,1} \right|. \label{eq:defZ}
\end{align}
Then we have
\begin{align}
    \E{Z} \leq \sqrt{8\pi} \frac{w(\set{K})}{\sqrt{mL}} 
\end{align}
and
\begin{align} \label{deviation}
    \P{Z > \frac{\sqrt{8\pi} w(\set{K})}{\sqrt{mL}} + u} \leq 2\exp \left( -\frac{mLu^2}{\pi d(\set{K})^2} \right) 
\end{align}
where $d(\set{K}) \coloneqq \max_{\x\in\set{K}} \pnorm{\x}{2}$.
\end{lemma}
\begin{Proof}
       Let $g\sim\set{N}(0,1)$ and note that $\E{\left|g\right|} = \sqrt{2/\pi}$. Then, we have
       \begin{align*}
           \E{\sum_{i=1}^m \sum_{l=1}^L \sqrt{\frac{\pi}{2Lm^2}} \left|\ip{ \al_i}{\x_{l}}\right|}
           = \sum_{i=1}^m \sum_{l=1}^L \sqrt{\frac{\pi}{2Lm^2}} \E{\left|g\right|}\pnorm{\x_{l}}{2}
           =\frac{\pnorm{\x}{2,1}}{\sqrt{L}}.
       \end{align*}
       Define now for $i \in [m], l \in [L]$ the random variables $\tl_i = \sqrt{\pi/(2Lm^2)} \left| \ip{ \al_i}{\x_{l}} \right|$, identically distributed independent copies $\thl_i$, and independent Rademacher variables $\eps_{i,l}$, i.e., $\P{\eps_{i,l}=1} = \P{\eps_{i,l}=-1}=1/2$. We obtain
\comm{       
       \begin{align}
           \E[\vartheta]{Z} &= \E[\vartheta]{\sup_{\x \in \set{K}} \left| \sum_{i=1}^m \sum_{l=1}^L (\tl_i - \E[\vartheta]{\tl_i}) \right| } \\
           &= \E[\vartheta]{\sup_{\x \in \set{K}} \left| \sum_{i=1}^m \sum_{l=1}^L \left( \tl_i - \E[\vartheta]{\tl_i} \right) - \E[\hat{\vartheta}]{\thl_i - \E[\hat{\vartheta}]{\thl_i}} \right| } \\
           &= \E[\vartheta]{\sup_{\x \in \set{K}} \left| \sum_{i=1}^m \sum_{l=1}^L \E[\hat{\vartheta}]{\tl_i - \thl_i} \right| } \\
           &\leqo{(i)} \E[\vartheta]{ \E[\hat{\vartheta}]{ \sup_{\x \in \set{K}} \left| \sum_{i=1}^m \sum_{l=1}^L \tl_i - \thl_i \right| }} \\
           &= \E[\vartheta]{ \E[\hat{\vartheta}]{ \sup_{\x \in \set{K}} \left| \sum_{i=1}^m \sum_{l=1}^L \eps_{i,l} \left(\tl_i - \thl_i\right) \right| }}\\
           &\leqo{(ii)} 2\E[\vartheta]{ \sup_{\x \in \set{K}} \left| \sum_{i=1}^m \sum_{l=1}^L \eps_{i,l} \tl_i \right| } \\
           &= 2\sqrt{\frac{\pi}{2Lm^2}} \E{ \sup_{\x \in \set{K}} \left| \sum_{i=1}^m \sum_{l=1}^L \eps_{i,l} \left| \ip{ \al_i}{\x_{l}} \right| \right| }\\
           &\leqo{(iii)} 4\sqrt{\frac{\pi}{2Lm^2}} \E{ \sup_{\x \in \set{K}} \left| \sum_{i=1}^m \sum_{l=1}^L \eps_{i,l} \ip{ \al_i}{\x_{l}} \right| } \\
           &= 4\sqrt{\frac{\pi}{2Lm^2}} \E{ \sup_{\x \in \set{K}} \left| \sum_{l=1}^L \ip{ \sum_{i=1}^m \eps_{i,l} \al_i}{\x_{l}} \right| }\\
           &= 4\sqrt{\frac{\pi}{2Lm^2}} \E{ \sup_{\x \in \set{K}} \left| \ip{ \sum_{i=1}^m \left( \eps_{i,1} ( \a_i^{(1)})^T,...,\eps_{i,L} (\a_i^{(L)})^T \right)^T}{\x} \right| } \\
           &\eqo{(iv)} 4\sqrt{\frac{\pi}{2Lm^2}} \E{ \sup_{\x \in \set{K}} \left| \ip{ \sqrt{m} \g}{\x} \right| } = \sqrt{8\pi} \frac{w(\set{K})}{\sqrt{mL}} 
        \label{eq:EUB}
       \end{align}
       }
       \begin{align}
           \E{Z} &= \E{\sup_{\x \in \set{K}} \left| \sum_{i=1}^m \sum_{l=1}^L (\tl_i - \E{\tl_i}) \right| } \\
           &= \E{\sup_{\x \in \set{K}} \left| \sum_{i=1}^m \sum_{l=1}^L \left( \tl_i - \E{\tl_i} \right) - \E{\thl_i - \E{\thl_i}} \right| } \\
           &= \E{\sup_{\x \in \set{K}} \left| \sum_{i=1}^m \sum_{l=1}^L \E{\tl_i - \thl_i} \right| } \\
           &\leqo{(i)} \E{ \E{ \sup_{\x \in \set{K}} \left| \sum_{i=1}^m \sum_{l=1}^L \tl_i - \thl_i \right| }} \\
           &= \E{ \E{ \sup_{\x \in \set{K}} \left| \sum_{i=1}^m \sum_{l=1}^L \eps_{i,l} \left(\tl_i - \thl_i\right) \right| }}\\
           &\leqo{(ii)} 2\E{ \sup_{\x \in \set{K}} \left| \sum_{i=1}^m \sum_{l=1}^L \eps_{i,l} \tl_i \right| } \\
           &= 2\sqrt{\frac{\pi}{2Lm^2}} \E{ \sup_{\x \in \set{K}} \left| \sum_{i=1}^m \sum_{l=1}^L \eps_{i,l} \left| \ip{ \al_i}{\x_{l}} \right| \right| }\\
           &\leqo{(iii)} 4\sqrt{\frac{\pi}{2Lm^2}} \E{ \sup_{\x \in \set{K}} \left| \sum_{i=1}^m \sum_{l=1}^L \eps_{i,l} \ip{ \al_i}{\x_{l}} \right| } \\
           &= 4\sqrt{\frac{\pi}{2Lm^2}} \E{ \sup_{\x \in \set{K}} \left| \sum_{l=1}^L \ip{ \sum_{i=1}^m \eps_{i,l} \al_i}{\x_{l}} \right| }\\
           &= 4\sqrt{\frac{\pi}{2Lm^2}} \E{ \sup_{\x \in \set{K}} \left| \ip{ \sum_{i=1}^m \left( \eps_{i,1} ( \a_i^{(1)})^T,...,\eps_{i,L} (\a_i^{(L)})^T \right)^T}{\x} \right| } \\
           &\eqo{(iv)} 4\sqrt{\frac{\pi}{2Lm^2}} \E{ \sup_{\x \in \set{K}} \left| \ip{ \sqrt{m} \g}{\x} \right| } = \sqrt{8\pi} \frac{w(\set{K})}{\sqrt{mL}} 
        \label{eq:EUB}
       \end{align}       
       where (i) follows from Jensen's inequality and (ii) from the triangle inequality, (iii) is a consequence of~\cite[Thm. 4.12]{ledoux:2013} and in (iv) we let  $\g \sim \set{N}(\0,\id_{mnL})$.To prove the deviation inequality \eqref{deviation} we will first show that $Z$, as defined in~\eqref{eq:defZ}, is Lipschitz continuous in $\A$. Consider two block diagonal matrices $\A,\Bmat$ as in~\eqref{eq:defA} and define the operator
    \begin{align}
        Z(\A) \coloneqq
        \sup_{\x\in\set{S}} \left| \sum_{i=1}^m \sum_{l=1}^L \sqrt{\frac{\pi}{2Lm^2}} \left|\ip{ \al_i}{\x_{l}}\right| - \frac{\pnorm{\x}{2,1}}{\sqrt{L}} \right|.
    \end{align}
    Then, we have
    \begin{align}
        &\hspace{-1em}| Z(\A) - Z(\Bmat) | \\
        &= \sup_{\x\in\set{K}} \left| \sum_{i=1}^m \sum_{l=1}^L  \sqrt{\frac{\pi}{2Lm^2}} \left|\ip{ \al_i}{\x_{l}}\right| - \frac{\pnorm{\x}{2,1}}{\sqrt{L}} \right| 
            - \sup_{\x\in\set{S}} \left| \sum_{i=1}^m \sum_{l=1}^L \sqrt{\frac{\pi}{2Lm^2}} \left|\ip{ \bl_i}{\x_{l}}\right| - \frac{\pnorm{\x}{2,1}}{\sqrt{L}} \right| \\
        &\leq  \sup_{\x\in\set{K}} \left\{ \left| \sum_{i=1}^m \sum_{l=1}^L \sqrt{\frac{\pi}{2Lm^2}} \left|\ip{ \al_i}{\x_{l}}\right| - \frac{\pnorm{\x}{2,1}}{\sqrt{L}} \right| 
            -  \left| \sum_{i=1}^m \sum_{l=1}^L \sqrt{\frac{\pi}{2Lm^2}} \left|\ip{ \bl_i}{\x_{l}}\right| - \frac{\pnorm{\x}{2,1}}{\sqrt{L}} \right| \right\}\\
        &\leq  \sup_{\x\in\set{K}} \left| \sum_{i=1}^m \sum_{l=1}^L \sqrt{\frac{\pi}{2Lm^2}} \left|\ip{ \al_i - \bl_i}{\x_{l}}\right|  \right| \\
        &\leq  \sup_{\x\in\set{K}} \sqrt{\frac{\pi}{2Lm^2}} \sum_{i=1}^m \sum_{l=1}^L   \pnorm{\al_i - \bl_i}{2} \pnorm{\x_{l}}{2}\\
        &\leq  \sup_{\x\in\set{K}} \sqrt{\frac{\pi}{2Lm^2}} \pnorm{\A - \Bmat}{F} \left( \sum_{i=1}^m \sum_{l=1}^L \pnorm{\x_{l}}{2}^2 \right)^{\frac{1}{2}} \\
        &\leq  \sqrt{\frac{\pi}{2Lm^2}} \sqrt{m} \pnorm{\A - \Bmat}{F} d(\set{K}) \\
        &= \frac{d(\set{K})}{\sqrt{mL}} \sqrt{\frac{\pi}{2}} \pnorm{\A - \Bmat}{F}
    \end{align}
    Hence, $Z(\cdot)$ is Lipschitz continuous with constant $\frac{d(\set{K})}{\sqrt{mL}} \sqrt{\frac{\pi}{2}}$. Using~\cite[Eq. (1.6)]{ledoux:2013}, we see that
    \begin{align}
        \P{|Z - \E{Z}| > u } \leq 2\exp\left( -\frac{2u^2 mL}{2\pi d(\set{K})^2} \right).
    \end{align}
    Thus, using~\eqref{eq:EUB}, we have
    \begin{align}
        \P{Z - \sqrt{8\pi} \frac{w(\set{K})}{\sqrt{mL}} > u } \leq \P{Z - \E{Z} > u } \leq \P{|Z - \E{Z}| > u } \leq 2\exp\left(- \frac{mL u^2}{\pi d(\set{K})^2 } \right),
    \end{align}
    which yields the claim.
\end{Proof}

\begin{Proof}[of Lemma \ref{lem:RIP}]
    The lemma is a direct consequence of Lemmas \ref{lem:GaussianWidth} and \ref{concentrationLemma}. Just choose $u=\delta/2$ and $mL \geq 8\pi (\delta/2)^{-2} w(\set{K}_{s,L} \cap \B(\0,1))^2 $ and note that by 
     Lemma \ref{lem:GaussianWidth}, we have $ w(\set{K}_{s,L} \cap \B(\0,1)) \ge w(\set{K}_{s,L} \cap \mathbb{S}^{nL-1})$. Then, with probability at least $1-2\exp\left(- mL \delta^2/(4\pi)\right)$, we have for all $\z \in \set{K}_{s,L} \cap \mathbb{S}^{nL-1}$
    \begin{align*}
    \left| \sqrt{\frac{\pi}{2Lm^2}} \pnorm{\A\z}{1} - \frac{\pnorm{\z}{2,1}}{\sqrt{L}}  \right| \le \sqrt{8\pi} \frac{w\left(\set{K}_{s,L} \cap \mathbb{S}^{nL-1}\right)}{\sqrt{mL}} + \frac{\delta}{2}  \le \delta.
    \end{align*}
    The statement follows for $\z \in \set{K}_{s,L}$ by multiplying $\pnorm{\z}{2}$ on both sides.
\end{Proof}


\subsection{Proof of Theorem \ref{thm:singleHT}} \label{sec:TheoremHT}
 Let us denote the set of nonzero rows of $\X$ by $ \supp(\X) = \supp(\x) = \set{T}$ for some $\set{T} \subset [n]$ with $|\set{T}| \le s$. For $\z = \vec(Z) \in \R^{nL}$ let $\z_\set{T} = \vec( \Z_\set{T} ) $ with $\Z_\set{T}$ being the matrix in which all rows not in $\set{T}$ are set to zero. The proof of Theorem \ref{thm:singleHT} follows the argument of \cite[Thm.\ 8]{foucart:2016} but relies on the assumption that all signals $\x_l$ share a common $\ell_2$-norm.
\begin{lemma} \label{lem:TechnicalLemma}
    If the operator $\theta \A$ satisfies the $\ell_1/\ell_{2,1}$-RIP on $\set{K}_{s,L}$, then all $\x \in \set{S}_{s,L}$ with $\pnorm{\x}{2} = 1$ satisfy
    \begin{align*}
        \pnorm{\left((\theta\A)^T \sign((\theta\A)\x)\right)_\set{T} - \x}{2}^2 \le 5\delta.
    \end{align*}
\end{lemma}
\begin{Proof}
    Define $\theta\b = \theta\A^T \sign (\A \x) \in \R^{nL}$ to be the backprojected quantized measurements. We then have
    \begin{align*}
        \pnorm{\left((\theta\A)^T \sign((\theta\A)\x)\right)_\set{T} - \x}{2}^2 = \pnorm{(\theta\b)_\set{T}}{2}^2 - 2\langle (\theta\b)_\set{T},\x \rangle + \pnorm{\x}{2}^2
    \end{align*}
    and
    \begin{align*}
        \pnorm{(\theta\b)_\set{T}}{2}^2 &= \langle (\theta\b)_\set{T},(\theta\b)_\set{T} \rangle 
        = \langle (\theta \A)^T \sign(\A\x), (\theta\b)_\set{T} \rangle  \\
        &= \langle \sign(\A\x), (\theta\A)(\theta\b)_\set{T} \rangle 
        \leq \pnorm{(\theta\A)(\theta\b)_\set{T}}{1}  \\
        & \leq \frac{\pnorm{(\theta\b)_\set{T}}{2,1}}{\sqrt{L}} + \delta \pnorm{(\theta\b)_\set{T}}{2}   
        \leq (1+\delta)\pnorm{(\theta\b)_\set{T}}{2}  
    \end{align*}
    Hence, we have $\pnorm{(\theta\b)_\set{T}}{2} \le 1+\delta$ and
    \begin{align*}
        \langle (\theta\b)_\set{T}, \x \rangle = \langle \sign(\A\x), (\theta\A) \x \rangle = \pnorm{(\theta\A)\x}{1}  \ge \frac{\pnorm{\x}{2,1}}{\sqrt{L}} - \delta \pnorm{\x}{2} = (1-\delta) 
    \end{align*}
    where we used that $\pnorm{\x}{2,1} = \sqrt{L} \pnorm{\x}{2} = \sqrt{L}$ by assumption. We can conclude that
    \begin{align*}
        \pnorm{\left((\theta\A)^T \sign((\theta\A)\x)\right)_\set{T} - \x}{2}^2 \le (1+\delta)^2 - 2 (1 - \delta) + 1 \le 5\delta.
    \end{align*}
\end{Proof}

\begin{Proof}[of Theorem \ref{thm:singleHT}]
	Choose $mL \gtrsim \delta^{-2} 2s(\log(en/(2s)) + L)$ such that by Lemma~\ref{lem:RIP}, $\theta\A$ satisfies the $\ell_1/\ell_{2,1}$-RIP on $\set{K}_{2s,L}$ with high probability.
    Let $\set{T} = \supp(\x)$ and $\hat{\set{T}} = \supp(\hat{\x})$ where $\hat{\x} = \mathbb{H}_s((\theta\A)^T\y)$. Note that $\hat{\x}$ is also the best $s$-row approximation of $((\theta\A)^T\y)_{\set{T} \cup \hat{\set{T}}}$. Hence,
    \begin{align*}
        \pnorm{\x - \hat{\x}}{2} &\le \pnorm{((\theta\A)^T\y)_{\set{T} \cup \hat{\set{T}}} - \hat{\x}}{2} + \pnorm{((\theta\A)^T\y)_{\set{T} \cup \hat{\set{T}}} - \x}{2} \\
        &\le 2\pnorm{((\theta\A)^T\y)_{\set{T} \cup \hat{\set{T}}} - \x}{2} \le 2\sqrt{5 \delta}. 
    \end{align*}
    where we applied Lemma \ref{lem:TechnicalLemma} for $\set{K}_{2s,L}$ in the last inequality (note that $|\set{T} \cup \hat{\set{T}}| \le 2s$).
\end{Proof}


\section{Numerical Experiments} \label{sec:NumericalExperiments}
In this section we illustrate numerically the theoretical results of Section \ref{sec:MainResults}. Recall that we propose to recover an unknown signal ensemble $\X \in \R^{n\times L}$ from its one-bit measurements $\Y \in \R^{m\times L}$ by a single hard-thresholding step which needs the measurements $\Y$, the block diagonal measurement matrix $\A$ and the sparsity level $s = |\supp(\X)|$. The simple approximation procedure is presented in Algorithm \ref{alg:singleHT}. We present two experiments which document 
the asymptotically linear dependence of $m = \set{O}(s)$ measurements per signal. In both experiments the block diagonal measurement matrix $\A$ has iid Gaussian entries and is scaled by $\theta = \sqrt{\pi / (2Lm^2)}$ as required in Lemma \ref{lem:RIP}. Signal ensembles $\X \in \R^{n\times L}$ with $|\supp(\X)| = s$ are created by first drawing some support set $\set{T} \subset [n]$ uniformly at random, then drawing the single entries as iid Gaussians of mean $0$ and variance $1$, and finally re-scaling all single signals $\x_l$, $l \in [L]$, to have unit norm.\\
    
\begin{algorithm}[t!]
	\caption{\textbf{:} \textbf{sHT}$(\y,\A,s)$} \label{alg:singleHT}
	\begin{algorithmic}[1]
		\Require{$\Y \in \{-1,1\}^{m\times L}$, $\A \in \R^{mL\times nL}$}
		\Statex
		\State $\hat{\x} \gets \mathbb{H}_s(\A^T\vec(\Y))$ \Comment{$\mathbb{H}_s$ is defined in \eqref{eq:singleHT}}
		\State $\hat{\X} \gets \textbf{reshape}(\x,n,L)$ \Comment{$\textbf{reshape}(\cdot)$ reverses $\vec(\cdot)$}
		\State
		\Return{$\hat{\X}$}
	\end{algorithmic}
\end{algorithm}

\begin{figure}[htb]
    \centering
%
%
\definecolor{mycolor1}{rgb}{0.00000,0.44700,0.74100}%
\definecolor{mycolor2}{rgb}{0.85000,0.32500,0.09800}%
\definecolor{mycolor3}{rgb}{0.92900,0.69400,0.12500}%
\definecolor{mycolor4}{rgb}{0.49400,0.18400,0.55600}%
\begin{tikzpicture}

\begin{axis}[%
width=9cm,
height=7cm,
at={(1.011111in,0.641667in)},
scale only axis,
xmode=log,
xmin=0.01,
xmax=3,
xminorticks=true,
xlabel={measurement rate $m/n$},
xmajorgrids,
xminorgrids,
ymode=log,
ymin=0,
ymax=4,
yminorticks=true,
ylabel={average $\pnorm{X- \hat{X}}{F} $},
ymajorgrids,
yminorgrids,
legend style={legend cell align=left,align=left,draw=white!15!black}
]
\addplot [color=mycolor1,solid,line width=2.0pt]
  table[row sep=crcr]{%
0.005	6.56904796880549\\
0.015	4.6825381741301\\
0.025	3.80826001855154\\
0.035	3.32052729592118\\
0.045	2.97800807079775\\
0.055	2.73365753500567\\
0.065	2.52967834820715\\
0.075	2.37799953709168\\
0.085	2.24678568705885\\
0.095	2.12739428701081\\
0.105	2.03232470825859\\
0.115	1.93078563489892\\
0.125	1.86548582149377\\
0.135	1.78211827893996\\
0.145	1.719712321561\\
0.155	1.66203886888046\\
0.165	1.58447014988938\\
0.175	1.5518916401584\\
0.185	1.49912774981861\\
0.195	1.47090340913701\\
0.205	1.43711776218692\\
0.215	1.38025360208857\\
0.225	1.35605151755177\\
0.235	1.32567929618131\\
0.245	1.29506480396745\\
0.25	1.28791036488397\\
0.275	1.20929744862733\\
0.31	1.14583328340472\\
0.34	1.07990174762666\\
0.38	1.02087307548701\\
0.42	0.952778647974287\\
0.465	0.897238696648448\\
0.515	0.848541884904624\\
0.57	0.797323733710691\\
0.635	0.753571286392525\\
0.705	0.711329464261188\\
0.78	0.663829295579413\\
0.865	0.618597175032471\\
0.96	0.579556153620727\\
1.065	0.542319457399807\\
1.18	0.506795516687844\\
1.31	0.47249310869104\\
1.455	0.441037671104784\\
1.61	0.414946376020865\\
1.79	0.393370286682814\\
1.985	0.364300835452554\\
2.2	0.344843264519651\\
2.44	0.320687439632166\\
2.705	0.297929982828377\\
3	0.287662503236342\\
};
\addlegendentry{$L = 1$};

\addplot [color=mycolor2,solid,line width=2.0pt]
  table[row sep=crcr]{%
0.005	5.58410460670625\\
0.015	3.98377488113837\\
0.025	3.30898631001163\\
0.035	2.87461430082554\\
0.045	2.58419765469802\\
0.055	2.36386537786739\\
0.065	2.17108605659144\\
0.075	2.03362560043623\\
0.085	1.92701227754647\\
0.095	1.8286118774421\\
0.105	1.75196913605957\\
0.115	1.66831476864978\\
0.125	1.57327532098189\\
0.135	1.52343823353542\\
0.145	1.47538381563417\\
0.155	1.41308780929026\\
0.165	1.36346476964293\\
0.175	1.3204847616\\
0.185	1.25804518943297\\
0.195	1.23887316949754\\
0.205	1.19717872017897\\
0.215	1.16229857524105\\
0.225	1.15338546846354\\
0.235	1.11020481321982\\
0.245	1.07563129099357\\
0.25	1.08586157216704\\
0.275	1.00278616385781\\
0.31	0.949097924543681\\
0.34	0.896915416906238\\
0.38	0.828910213377369\\
0.42	0.76009196707777\\
0.465	0.726568365558056\\
0.515	0.661251767822911\\
0.57	0.614808298364823\\
0.635	0.585313379083515\\
0.705	0.540597155438207\\
0.78	0.506370807504775\\
0.865	0.466594042314659\\
0.96	0.437743325442995\\
1.065	0.401564751083986\\
1.18	0.381743465853744\\
1.31	0.340678295935754\\
1.455	0.318399939348002\\
1.61	0.30129142188554\\
1.79	0.272976219861984\\
1.985	0.256313921405036\\
2.2	0.235415049084473\\
2.44	0.214493429795564\\
2.705	0.205782238630038\\
3	0.191654133267446\\
};
\addlegendentry{$L = 2$};

\addplot [color=mycolor3,solid,line width=2.0pt]
  table[row sep=crcr]{%
0.005	4.63291040669785\\
0.015	3.35264014462228\\
0.025	2.77461000519775\\
0.035	2.42222233331292\\
0.045	2.16809694452549\\
0.055	1.97470612883522\\
0.065	1.83699467408758\\
0.075	1.7184182391799\\
0.085	1.58923230649772\\
0.095	1.52046672770967\\
0.105	1.42736348483806\\
0.115	1.34937090092704\\
0.125	1.29196557090939\\
0.135	1.22160929095067\\
0.145	1.17862472039805\\
0.155	1.13388902997462\\
0.165	1.07635182120972\\
0.175	1.03494962187061\\
0.185	0.998244315206413\\
0.195	0.970189902209016\\
0.205	0.928406590012997\\
0.215	0.891008064918636\\
0.225	0.864035117702277\\
0.235	0.8430796335827\\
0.245	0.815584359093528\\
0.25	0.82718023194547\\
0.275	0.736936526038378\\
0.31	0.695987593017945\\
0.34	0.648986470583865\\
0.38	0.585517117710068\\
0.42	0.529392629731084\\
0.465	0.498671467464952\\
0.515	0.456774841192215\\
0.57	0.426181255667461\\
0.635	0.391489097336514\\
0.705	0.356568586107544\\
0.78	0.336388266201235\\
0.865	0.307762559547234\\
0.96	0.282787383246484\\
1.065	0.270019563114004\\
1.18	0.251045874830245\\
1.31	0.238986673951856\\
1.455	0.223589321210784\\
1.61	0.20940262046248\\
1.79	0.197763541539425\\
1.985	0.187631317067989\\
2.2	0.176741815666964\\
2.44	0.166989101148092\\
2.705	0.157504422750711\\
3	0.1495049176589\\
};
\addlegendentry{$L = 5$};

\addplot [color=mycolor4,solid,line width=2.0pt]
  table[row sep=crcr]{%
0.005	3.80721041361774\\
0.015	2.75196993549576\\
0.025	2.27343959728216\\
0.035	1.97249624052223\\
0.045	1.7439019171987\\
0.055	1.56856816096645\\
0.065	1.41765876098753\\
0.075	1.28745311868121\\
0.085	1.18168483371855\\
0.095	1.09418039333796\\
0.105	1.00706024493754\\
0.115	0.952069388643124\\
0.125	0.886406944658128\\
0.135	0.83035094757919\\
0.145	0.78308070737759\\
0.155	0.746064198582969\\
0.165	0.70417458207177\\
0.175	0.673208278103342\\
0.185	0.641241593177266\\
0.195	0.61999506268706\\
0.205	0.602252477338906\\
0.215	0.583788618183311\\
0.225	0.566678003691427\\
0.235	0.544607963708499\\
0.245	0.529257783451583\\
0.25	0.535315918754654\\
0.275	0.499015206307549\\
0.31	0.473799821700612\\
0.34	0.448453830879615\\
0.38	0.424076155249264\\
0.42	0.39784203879862\\
0.465	0.380355957633352\\
0.515	0.363402172532261\\
0.57	0.343400442637556\\
0.635	0.32515175647421\\
0.705	0.31013007938226\\
0.78	0.295415967775695\\
0.865	0.279702959294121\\
0.96	0.263836410342657\\
1.065	0.253558711319084\\
1.18	0.239488101684103\\
1.31	0.229018662146136\\
1.455	0.217303930519964\\
1.61	0.206698151289066\\
1.79	0.194952724308146\\
1.985	0.184372876907782\\
2.2	0.175471681372106\\
2.44	0.167071920998202\\
2.705	0.158447340232295\\
3	0.150931342046789\\
};
\addlegendentry{$L = 20$};

\end{axis}
\end{tikzpicture}%
    \caption{Simulated error $\| \X - \hat{\X} \|_F$ averaged over 500 experiments for $s=5$ and $n=100$.}
    \label{fig:Lcomparison}
\end{figure}
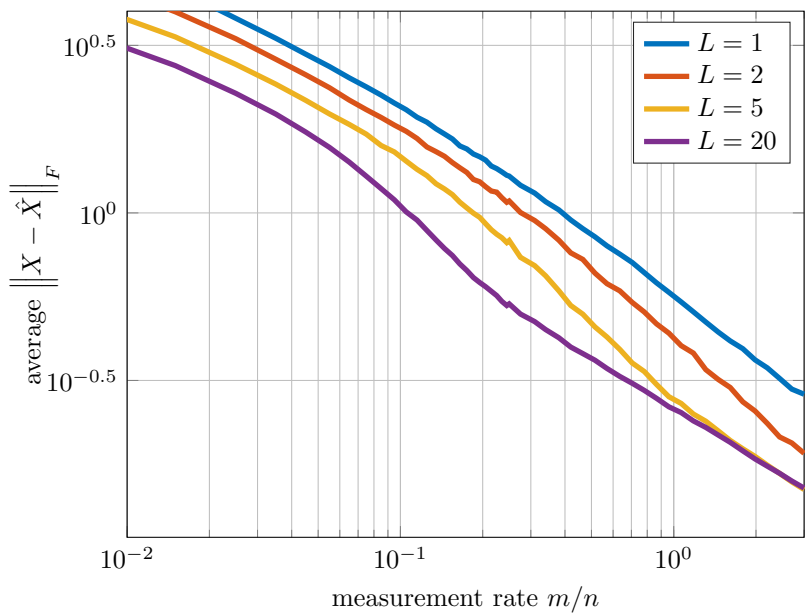

In the first experiment we approximate $500$ randomly drawn signal ensembles $\X \in \R^{n\times L}$ of signal dimension $n = 100$, ensemble size $L = 1,2,5,20$, and support size $s = 5$ from their one-bit measurements $\Y \in \R^{m \times L}$. Figure \ref{fig:Lcomparison} depicts the in average obtained approximation error $\| \X - \hat{\X} \|_F$ in Frobenius norm over the measurement rate $r = m/n$. One clearly observes an improvement for larger ensembles. In particular, there is a sharper transition from no recovery to practical approximation.\\

\begin{figure}[htb]
    \centering
    \begin{minipage}{0.55\textwidth}
	\input{distributed_transition.tex}
\end{minipage}
\hfill
\hspace{2em}
\begin{minipage}{0.1\textwidth}
	\includegraphics[width=0.55\textwidth]{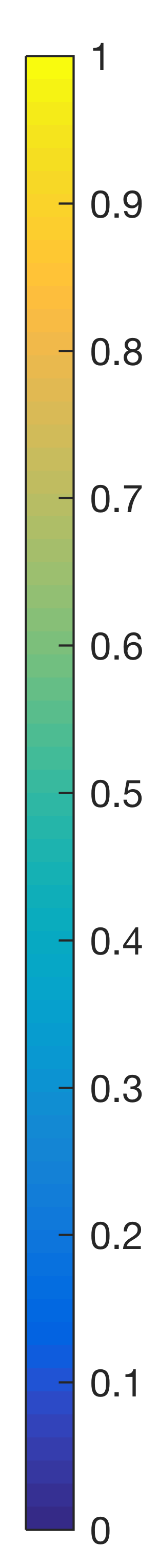}
\end{minipage}
    \caption{Simulated error $\| \X - \hat{\X} \|_F$ averaged over 500 experiments with $n=100$. The red contour lines correspond to $\| \X - \hat{\X} \|_F=2/3$.}
    \label{fig:phase_diag}
\end{figure}

The second experiment (see Figure \ref{fig:phase_diag}) illustrates the dependence of $m$ and $s$. We again approximate $500$ randomly drawn signal ensembles $\X \in \R^{n\times L}$ of signal dimension $n = 100$ and ensemble size $L = 1,2,5,20$ from their one-bit measurements $\Y \in \R^{m \times L}$. This time the support size of $\X$ varies from $s=1$ to $s=50$ while the measurement rate $r = m/n$ ranges from $r = 0.01$ up to $r = 3$. The average approximation error $\| \X - \hat{\X} \|_F$ is plotted in color while a selected error level is highlighted. When comparing the different choices of $L$, the linear dependence of $m$ on $s$ for $L = 20$ and fixed error levels is clearly visible and different from the $s\ln (en/s)$ behavior for $L = 1$.

The reader might notice that the measurement rate does not behave linearly in the plots $L=2$ and $L=5$ for $s/n \ge e^{1-L}$ which corresponds to the case $L \ge \log(en/s)$ for which we claimed $\mathcal{O}(s)$ behaviour in Remark \ref{rem:RIP}. This, however, is no contradiction as for the $\mathcal{O}(s)$ argument it suffices to bound $(\log(en/s) + L) /L \leq 2$. In the numerical experiments with fixed $L$ we observe the transition from $(\log(en/s) + L) /L \approx 2$ for small values of $s/n$ (corresponding to large values of $\log(en/s)$) to $(\log(en/s) + L) /L \approx 1$ for large values of $s/n$ (corresponding to small values of $\log(en/s)$) causing a non-linear shape as long as $L$ is not clearly dominating (cf.\ $L = 20$).


\section{Conclusion} \label{sec:Conclusion}
In the presented work we examined how the two concepts of heavily quantized measurements and distributed compressed sensing can be combined. We showed that a single hard thresholding step enables uniform joint approximation of several signals sharing a common support for $m = \set{O}(s)$ measurements per signal. We see two possible directions of future research. First, sophisticated alternatives to a single hard-thresholding step have been proposed (see \red{\cite{kafle2016decentralized}}) which numerically give a smaller approximation error. It would be nice to extend the theory also to these methods. Second, relaxing the quantization level to multi-bit quantizers is desirable as it should decrease the approximation error (cf.\ \red{\cite{jacques:2013,jacques2016error}}) and thus bridges the wide performance gap between unquantized measurements and one-bit measurements.

\section*{Acknowledgements} \label{sec:Acknowledgements}
We would like to thank Laurent Jacques for helpful discussion that led to a significant improvement of the results presented in this work.
Further, we would like to thank Gerhard Kramer for valuable comments that improved the presentation of this work.


%








\bibliography{library}

\begin{thebibliography}{10}
\providecommand{\url}[1]{#1}
\csname url@samestyle\endcsname
\providecommand{\newblock}{\relax}
\providecommand{\bibinfo}[2]{#2}
\providecommand{\BIBentrySTDinterwordspacing}{\spaceskip=0pt\relax}
\providecommand{\BIBentryALTinterwordstretchfactor}{4}
\providecommand{\BIBentryALTinterwordspacing}{\spaceskip=\fontdimen2\font plus
\BIBentryALTinterwordstretchfactor\fontdimen3\font minus
  \fontdimen4\font\relax}
\providecommand{\BIBforeignlanguage}[2]{{%
\expandafter\ifx\csname l@#1\endcsname\relax
\typeout{** WARNING: IEEEtranS.bst: No hyphenation pattern has been}%
\typeout{** loaded for the language `#1'. Using the pattern for}%
\typeout{** the default language instead.}%
\else
\language=\csname l@#1\endcsname
\fi
#2}}
\providecommand{\BIBdecl}{\relax}
\BIBdecl

\bibitem{baraniuk2017exponential}
R.~G. Baraniuk, S.~Foucart, D.~Needell, Y.~Plan, and M.~Wootters, ``Exponential
  decay of reconstruction error from binary measurements of sparse signals,''
  \emph{IEEE Transactions on Information Theory}, vol.~63, no.~6, pp.
  3368--3385, 2017.

\bibitem{baron2009distributed}
D.~Baron, M.~F. Duarte, M.~B. Wakin, S.~Sarvotham, and R.~G. Baraniuk,
  ``Distributed compressive sensing,'' \emph{arXiv preprint arXiv:0901.3403},
  2009.

\bibitem{boufounos:2008}
P.~T. Boufounos and R.~G. Baraniuk, ``1-bit compressive sensing,'' in
  \emph{42nd Annual Conference on Information Sciences and Systems (CISS)},
  2008, pp. 16--21.

\bibitem{candes2011tight}
E.~J. Cand\`es and Y.~Plan, ``Tight oracle inequalities for low-rank matrix
  recovery from a minimal number of noisy random measurements,'' \emph{IEEE
  Transactions on Information Theory}, vol.~57, no.~4, pp. 2342--2359, 2011.

\bibitem{candes2006robust}
E.~J. Cand{\`e}s, J.~Romberg, and T.~Tao, ``Robust uncertainty principles:
  Exact signal reconstruction from highly incomplete frequency information,''
  \emph{IEEE Transactions on Information Theory}, vol.~52, no.~2, pp. 489--509,
  2006.

\bibitem{candes2006stable}
E.~J. Cand\`es, J.~K. Romberg, and T.~Tao, ``Stable signal recovery from
  incomplete and inaccurate measurements,'' \emph{Communications on pure and
  applied mathematics}, vol.~59, no.~8, pp. 1207--1223, 2006.

\bibitem{candes2006near}
E.~J. Cand\`es and T.~Tao, ``Near-optimal signal recovery from random
  projections: Universal encoding strategies?'' \emph{IEEE Transactions on
  Information Theory}, vol.~52, no.~12, pp. 5406--5425, 2006.

\bibitem{donoho2006compressed}
D.~L. Donoho, ``Compressed sensing,'' \emph{IEEE Transactions on Information
  Theory}, vol.~52, no.~4, pp. 1289--1306, 2006.

\bibitem{duarte2005joint}
M.~F. Duarte, S.~Sarvotham, M.~B. Wakin, D.~Baron, and R.~G. Baraniuk, ``Joint
  sparsity models for distributed compressed sensing,'' in \emph{Proceedings of
  the Workshop on Signal Processing with Adaptative Sparse Structured
  Representations}.\hskip 1em plus 0.5em minus 0.4em\relax IEEE, 2005.

\bibitem{eftekhari2015restricted}
A.~Eftekhari, H.~L. Yap, C.~J. Rozell, and M.~B. Wakin, ``The restricted
  isometry property for random block diagonal matrices,'' \emph{Applied and
  Computational Harmonic Analysis}, vol.~38, no.~1, pp. 1--31, 2015.

\bibitem{eldar:2009}
Y.~C. Eldar and M.~Mishali, ``Robust recovery of signals from a structured
  union of subspaces,'' \emph{{I}{E}{E}{E} {T}rans. {I}nf. {T}heory}, vol.~55,
  no.~11, pp. 5302 -- 5316, 2009.

\bibitem{eldar:2010}
Y.~C. Eldar and H.~Rauhut, ``Average case analysis of multichannel sparse
  recovery using convex relaxation,'' \emph{{I}{E}{E}{E} {T}rans. {I}nf.
  {T}heory}, vol.~56, no.~1, pp. 505--519, 2010.

\bibitem{foucart:2016}
S.~Foucart, ``Flavors of compressive sensing,'' in \emph{Int. Conf.
  Approximation Theory}.\hskip 1em plus 0.5em minus 0.4em\relax Springer, 2016.

\bibitem{foucart:2013}
S.~Foucart and H.~Rauhut, \emph{A Mathematical Introduction to Compressive
  Sensing}.\hskip 1em plus 0.5em minus 0.4em\relax Birkh\"auser Basel, 2013.

\bibitem{genzel2017recovering}
M.~Genzel and P.~Jung, ``Recovering structured data from superimposed
  non-linear measurements,'' \emph{arXiv preprint arXiv:1708.07451}, 2017.

\bibitem{gupta2015joint}
V.~Gupta, B.~Kailkhura, T.~Wimalajeewa, S.~Liu, and P.~K. Varshney, ``Joint
  sparsity pattern recovery with 1-bit compressive sensing in sensor
  networks,'' in \emph{Signals, Systems and Computers, 2015 49th Asilomar
  Conference on}.\hskip 1em plus 0.5em minus 0.4em\relax IEEE, 2015, pp.
  1472--1476.

\bibitem{jacques2016error}
L.~Jacques, ``Error decay of (almost) consistent signal estimations from
  quantized gaussian random projections,'' \emph{IEEE Transactions on
  Information Theory}, vol.~62, no.~8, pp. 4696--4709, 2016.

\bibitem{jacques:2013}
L.~Jacques, K.~Degraux, and C.~D. Vleeschouwer, ``Quantized iterative hard
  thresholding: Bridging 1-bit and high-resolution quantized compressed
  sensing,'' in \emph{Int. Conf. Sampling Theory and Applications}, 2013.

\bibitem{kafle2016decentralized}
S.~Kafle, B.~Kailkhura, T.~Wimalajeewa, and P.~K. Varshney, ``Decentralized
  joint sparsity pattern recovery using 1-bit compressive sensing,'' in
  \emph{Signal and Information Processing (GlobalSIP), 2016 IEEE Global
  Conference on}.\hskip 1em plus 0.5em minus 0.4em\relax IEEE, 2016, pp.
  1354--1358.

\bibitem{knudson2016one}
K.~Knudson, R.~Saab, and R.~Ward, ``One-bit compressive sensing with norm
  estimation,'' \emph{IEEE Transactions on Information Theory}, vol.~62, no.~5,
  pp. 2748--2758, 2016.

\bibitem{ledoux:2013}
M.~Ledoux and M.~Talagrand, \emph{Probability in Banach Spaces: isoperimetry
  and processes}.\hskip 1em plus 0.5em minus 0.4em\relax Springer Verlag, 2002.

\bibitem{plan:2013}
Y.~Plan and R.~Vershynin, ``One-bit compressed sensing by linear programming,''
  \emph{Comm. Pure Appl. Math.}, vol.~66, pp. 1275 -- 1297, 2013.

\bibitem{plan:2013a}
------, ``Robust 1-bit compressed sensing and sparse logistic regression: A
  convex programming approach,'' \emph{IEEE Transactions on Information
  Theory}, vol.~59, no.~1, pp. 482 -- 494, 2013.

\bibitem{plan:2014}
------, ``Dimension reduction by random hyperplane tesselation,''
  \emph{Discrete \& Computational Geometry}, vol.~51, no.~2, pp. 438--461,
  2014.

\bibitem{rao:2014}
X.~Rao and V.~K.~N. Lau, ``Distributed compressive {CSIT} estimation and
  feedback for {FDD} multi-user massive {MIMO} systems,'' \emph{IEEE
  Transaction on Signal Processing}, vol.~62, no.~12, pp. 3261 -- 3271, Jun
  2014.

\bibitem{rudelson2008sparse}
M.~Rudelson and R.~Vershynin, ``On sparse reconstruction from fourier and
  {G}aussian measurements,'' \emph{Communications on Pure and Applied
  Mathematics}, vol.~61, no.~8, pp. 1025--1045, 2008.

\bibitem{schechtman2006two}
G.~Schechtman, ``Two observations regarding embedding subsets of euclidean
  spaces in normed spaces,'' \emph{Advances in Mathematics}, vol. 200, no.~1,
  pp. 125--135, 2006.

\bibitem{sundman2013methods}
D.~Sundman, S.~Chatterjee, and M.~Skoglund, ``Methods for distributed
  compressed sensing,'' \emph{Journal of Sensor and Actuator Networks}, vol.~3,
  no.~1, pp. 1--25, 2013.

\bibitem{talagrand2014upper}
M.~Talagrand, \emph{Upper and lower bounds for stochastic processes: modern
  methods and classical problems}.\hskip 1em plus 0.5em minus 0.4em\relax
  Springer Science \& Business Media, 2014, vol.~60.

\bibitem{tian2014distributed}
Y.~Tian, W.~Xu, Y.~Wang, and H.~Yang, ``A distributed compressed sensing scheme
  based on one-bit quantization,'' in \emph{Vehicular Technology Conference
  (VTC Spring), 2014 IEEE 79th}.\hskip 1em plus 0.5em minus 0.4em\relax IEEE,
  2014, pp. 1--6.

\bibitem{wu:2014}
Y.~Wu, Y.-J. Zhu, Q.-Y. Tang, C.~Zhu, W.~Liu, R.-B. Dai, X.~Liu, E.~X. Wu,
  L.~Ying, and D.~Liang, ``Accelerated {MR} diffusion tensor imaging using
  distributed compressed sensing,'' \emph{Magnetic Resonance in Medicine},
  vol.~71, no.~2, pp. 764 -- 772, 2014.

\end{thebibliography}
\bibliographystyle{IEEEtranS}
\end{document}